\documentclass[conference]{IEEEtran}
\IEEEoverridecommandlockouts

\usepackage{cite}
\usepackage{amsmath,amssymb,amsfonts}
\usepackage{graphicx}
\usepackage{textcomp}
\usepackage{xcolor}
\usepackage[utf8]{inputenc}
\usepackage{booktabs}
\usepackage{listings}
\usepackage{xcolor}
\usepackage{hyperref}
\definecolor{lightblue}{RGB}{173, 216, 230}  
\definecolor{lightgreen}{RGB}{204, 235, 197} 
\definecolor{lightgray}{RGB}{220, 220, 220} 
\definecolor{light-gray}{gray}{0.97}
\usepackage[table]{xcolor}
\usepackage{amsthm}
\newtheorem{example}{Example}
\usepackage[linesnumbered,ruled,vlined]{algorithm2e}
\usepackage{amsmath}

\usepackage{fp} 

\usepackage{hyperref}
\usepackage{hyperxmp}
\usepackage{xspace}
\usepackage[utf8]{inputenc}
\usepackage{balance}
\usepackage{amsmath}  
\usepackage{booktabs}
\usepackage{caption}
\usepackage{colortbl}
\usepackage{xcolor}    
\usepackage{graphicx}
\usepackage{tikz}
\usetikzlibrary{patterns}
\usepackage{siunitx}   
\usepackage{multirow}
\usepackage{footmisc}
\usepackage{listings}
\usepackage{commath}
\usepackage{epstopdf}
\usepackage{url}
\usepackage{float}
\usepackage{subcaption}
\usepackage{mathrsfs}
\usepackage{bbm}
\usepackage{fancyvrb}
\usepackage[linesnumbered,ruled,vlined]{algorithm2e}
\usepackage{algpseudocode}
\usepackage{mathtools}
\usepackage{enumitem}
\usepackage{array}
\usepackage{arydshln}
\usepackage{cleveref}

\usepackage{float}
\newcommand{\mcs}{\texttt{MCS}\xspace}
\renewcommand{\ss}{\texttt{SS}\xspace}
\newcommand{\rss}{\texttt{RSS}\xspace}
\newcommand{\tpch}{\texttt{TPC-H}\xspace}
\newcommand{\arss}{\texttt{ARSS}\xspace}
\newcommand{\ass}{\texttt{ASS}\xspace}

\newcommand{\nrel}{\texttt{\#rel}\xspace}
\newcommand{\nans}{\texttt{\#answ}\xspace}

\newcommand{\ncolmn}{\texttt{\#colmn}\xspace}
\newcommand{\rtime}{\texttt{time}\xspace}
\newcommand{\ptime}{\texttt{ptime}\xspace}
\newcommand{\psize}{\texttt{psize}\xspace}
\newcommand{\q}[1]{\ensuremath{\texttt{Q}_\texttt{#1}}}
\newcommand{\perm}{\texttt{Perm}\xspace}
\newcommand{\mre}{\texttt{MRE}\xspace}

\newcommand{\timeinseconds}[1]{%
  \FPeval{\result}{round(#1/1000,2)}%
  \result%
}

\newcommand{\formattime}[3]{%
  \pgfmathparse{round(#1*10^#3)/10^#3}%
  \edef\formattedtime{\pgfmathprintnumber[fixed,precision=#3]{\pgfmathresult}}%
  \formattedtime~#2%
}

\usetikzlibrary{shapes, arrows, positioning}

\definecolor{mygreen}{rgb}{0,0.6,0}
\definecolor{myred}{rgb}{0.6,0,0}
\definecolor{mygray}{rgb}{0.5,0.5,0.5}
\definecolor{mymauve}{rgb}{0.58,0,0.82}
\definecolor{myblue}{rgb}{0,0,1}

\usepackage{listings}

\definecolor{light-gray}{gray}{0.97}

\lstdefinestyle{promptstyle}{
    basicstyle=\rmfamily\footnotesize,
    backgroundcolor=\color{light-gray},
    keywordstyle=\color{blue},
    stringstyle=\color{red},
    commentstyle=\color{gray},
    showstringspaces=false,
    breaklines=true,
    tabsize=4
}

\usepackage{tikz}
\usetikzlibrary{positioning, arrows.meta}

\definecolor{highlight}{rgb}{0.9, 0.9, 0.5}

\DeclareMathAlphabet\mathbfcal{OMS}{cmsy}{b}{n}

\definecolor{ReviewerOneColor}{RGB}{240, 110, 0}   
\definecolor{ReviewerTwoColor}{RGB}{0, 139, 109}   
\definecolor{ReviewerThreeColor}{RGB}{45, 85, 205}
\definecolor{MetaReviewerColor}{RGB}{142, 69, 133} 

\definecolor{darkred}{rgb}{0.55, 0.0, 0.0}

\definecolor{Teal}{HTML}{1B9E77}    
\definecolor{SkyBlue}{HTML}{56B4E9} 
\definecolor{Vermillion}{HTML}{D55E00} 

\usepackage{amsthm}

\usepackage{colortbl}  
\usepackage{xcolor}    

\definecolor{lightblue}{RGB}{173, 216, 230}  
\definecolor{lightgreen}{RGB}{204, 235, 197} 
\definecolor{lightgray}{RGB}{220, 220, 220} 
\definecolor{lightred}{RGB}{255, 204, 203}

\newcommand{\boxtheorem}{\hfill $\blacksquare$\vspace{2mm}}

\newcommand{\ignore}[1]{}

\definecolor{black}{rgb}{0,0,0}
\definecolor{grey}{rgb}{0.8,0.8,0.8}
\definecolor{red}{rgb}{1,0,0}
\definecolor{green}{rgb}{0,1,0}
\definecolor{darkgreen}{rgb}{0,0.5,0}
\definecolor{darkpurple}{rgb}{0.5,0,0.5}
\definecolor{darkdarkpurple}{rgb}{0.3,0,0.3}
\definecolor{blue}{rgb}{0,0,1}
\definecolor{shadegreen}{rgb}{0.95,1,0.95}
\definecolor{shadeblue}{rgb}{0.95,0.95,1}
\definecolor{shadered}{rgb}{1,0.85,0.85}
\definecolor{shadegrey}{rgb}{0.85,0.85,0.85}
\definecolor{oddRowGrey}{rgb}{0.80,0.80,0.80}
\definecolor{evenRowGrey}{rgb}{0.85,0.85,0.85}
\definecolor{lightpurple}{rgb}{0.88,1.0,1.0}

\newcommand{\RNum}[1]{\uppercase\expandafter{\romannumeral #1\relax}}

\newtheorem{defn}{Definition}[section]

\newtheorem{col}[defn]{Colollary}

\newcommand{\proj}[1]{{\Pi}}
\newcommand{\sel}[1]{{\sigma}}

\newcommand{\cut}[1]{}
\newcommand{\eat}[1]{}

\SetKwInput{KwInput}{Input}
\SetKwInput{KwOutput}{Output}

\ExplSyntaxOn
  \cs_new_eq:NN \calc \fp_eval:n
\ExplSyntaxOff

\definecolor{cbgreen}{RGB}{0,158,115} 
\definecolor{cbred}{RGB}{213,94,0} 

\definecolor{DodgerBlue}{RGB}{30, 144, 255}

\newcommand{\permt}{\pi}

\newcommand{\Endg}[2]{E_#2(#1,D)}
\newcommand{\EndgR}[1]{E(#1, D)}

\newcommand{\setting}[2]{\texttt{#2-}\q{#1}}

\usepackage{xfp} 

\newcommand{\qtime}{\texttt{q-time}\xspace}
\newcommand{\ttime}{\texttt{t-time}\xspace}

\newcommand{\hl}[1]{\textbf{#1}}      
\newcommand{\shl}[1]{\underline{#1}}  

\lstdefinelanguage{SQL}{
  morekeywords={SELECT,FROM,WHERE,AND,OR,NOT,JOIN,ON,AS,IN,EXISTS,GROUP, BY, HAVING},
  sensitive=false,
  morecomment=[l]--,
  morestring=[b]',
}
\lstdefinestyle{promptstyle}{
    basicstyle=\rmfamily\small,
    backgroundcolor=\color{light-gray},
    keywordstyle=\color{blue},
    stringstyle=\color{red},
    commentstyle=\color{gray},
    showstringspaces=false,
    breaklines=true,
    tabsize=4
}
\def\BibTeX{{\rm B\kern-.05em{\sc i\kern-.025em b}\kern-.08em
    T\kern-.1667em\lower.7ex\hbox{E}\kern-.125emX}}
\begin{document}


\title{Relation-Stratified Sampling for Shapley Values Estimation in Relational Databases}

\author{
\IEEEauthorblockN{Amirhossein Alizad}
\IEEEauthorblockA{\textit{Department of Computer Science} \\
\textit{Western University}\\
London, Ontario, Canada \\
aaliza8@uwo.ca}
\and
\IEEEauthorblockN{Mostafa Milani}
\IEEEauthorblockA{\textit{Department of Computer Science} \\
\textit{Western University}\\
London, Ontario, Canada \\
mostafa.milani@uwo.ca}
}

\maketitle

\begin{abstract}
Shapley-like values, including the Shapley and Banzhaf values, provide a principled way to quantify how individual tuples contribute to a query result. Their exact computation, however, is intractable because it requires aggregating marginal contributions over exponentially many permutations or subsets. While sampling-based estimators have been studied in cooperative game theory, their direct use for relational query answering remains underexplored and often ignores the structure of schemas and joins.

We study tuple-level attribution for relational queries through sampling and introduce \emph{Relation-Stratified Sampling} (\rss). Instead of stratifying coalitions only by size, \rss partitions the sample space by a relation-wise count vector that records how many tuples are drawn from each relation. This join-aware stratification concentrates samples on structurally valid and informative coalitions and avoids strata that cannot satisfy query conditions. We further develop an adaptive variant, \arss, that reallocates budget across strata using variance estimates obtained during sampling, improving estimator efficiency without increasing the total number of samples. We analyze these estimators, describe a practical implementation that reuses compiled views to reduce per-sample query cost, and evaluate them on \tpch workloads.

Across diverse queries with multi-relation joins and aggregates, \rss and \arss consistently outperform classical Monte Carlo (\mcs) and size-based Stratified Sampling (\ss), yielding lower error and variance with fewer samples. An ablation shows that relation-aware stratification and adaptive allocation contribute complementary gains, making \arss a simple, effective, and anytime estimator for database-centric Shapley attribution.
\end{abstract}

\begin{IEEEkeywords}
database, query, Shapley, estimation, sampling, approximation, Monte Carlo, Stratified, Relation Stratified
\end{IEEEkeywords}

\section{Introduction}
\label{sec:intro}

Shapley-like values, including the Shapley value~\cite{shapley1953value} and related indices such as the Banzhaf index~\cite{banzhaf1965weighted}, provide a principled foundation for attributing the outcome of a computation to its individual inputs based on their marginal contributions. Originally proposed in cooperative game theory, these values are widely appreciated for their fairness properties—efficiency, symmetry, additivity, and the null player condition. Over time, their use has expanded beyond economics and game theory~\cite{roth1988shapley} to artificial intelligence and machine learning~\cite{chalkiadakis2011computational,lundberg2017unified}, where they are employed for attribution, explanation, and fairness analysis.

In data management, Shapley-based reasoning treats database tuples as “players’’ and the value of a coalition as the output of a query. This enables analysis of how individual tuples influence query results, supporting tasks such as query explanation, provenance summarization~\cite{alomeir2023summarizing,alomeir2020pastwatch}, contributor ranking, and data cleaning. The same axiomatic rigor that makes Shapley values attractive also makes them computationally challenging: computing an exact value requires averaging a tuple’s marginal contribution over all subsets (or permutations) of the other tuples, which is exponential in the number of contributing records. In databases, each marginal contribution typically entails at least one query execution on a sub-instance, quickly rendering naïve computation infeasible.

\begin{table}[t!]
\centering
\resizebox{0.95\linewidth}{!}{
\begin{tabular}{lcccc}
\toprule
\textbf{LID} & \textbf{OrderKey} & \textbf{ExtPrice} & \textbf{Discount} & \textbf{ShipDate} \\
\midrule
\rowcolor{lightgray} $\ell_1$ & 23417 & 500 & 0.10 & 1998-04-21 \\
\rowcolor{lightgray} $\ell_2$ & 23417 & 600 & 0.01 & 1998-04-16 \\
\rowcolor{lightgray} $\ell_3$ & 23417 & 700 & 0.06 & 1998-04-06 \\
\rowcolor{lightgray} $\ell_4$ & 23417 & 650 & 0.05 & 1998-03-25 \\
$\ell_5$ & 23110 & 820 & 0.04 & 1998-02-14 \\
$\ell_6$ & 23110 & 560 & 0.03 & 1998-02-17 \\
$\ell_7$ & 22789 & 400 & 0.07 & 1998-01-11 \\
$\ell_8$ & 22789 & 720 & 0.06 & 1998-01-15 \\
\bottomrule
\end{tabular}}
\caption{\texttt{lineitem} Relation}
\label{tab:lineitem}
\end{table}

\begin{table}[t!]
\centering
\begin{tabular}{lcccc}
\toprule
\textbf{CID} & \textbf{CustKey} & \textbf{Name} & \textbf{AcctBal} & \textbf{MktSegment} \\
\midrule
\rowcolor{lightgray} $c_1$ & 1456 & Cust1456 & 6800 & AUTO \\
$c_2$ & 3125 & Cust3125 & 4300 & MACHINERY \\
\bottomrule
\end{tabular}
\caption{\texttt{customer} Relation}
\label{tab:customer}
\end{table}

\begin{table}[t!]
\centering
\begin{tabular}{lcccc}
\toprule
\textbf{OID} & \textbf{OrderKey} & \textbf{CustKey} & \textbf{OrderDate} & \textbf{ShipPrio} \\
\midrule
\rowcolor{lightgray} $o_1$ & 23417 & 1456 & 1997-12-21 & 0 \\
$o_2$ & 23110 & 3125 & 1998-01-05 & 1 \\
$o_3$ & 22789 & 3125 & 1998-01-07 & 0 \\
\bottomrule
\end{tabular}
\caption{\texttt{orders} Relation}
\label{tab:orders}
\end{table}

\begin{example} \em \label{ex:tpch-intro}
Consider a simplified version of the \tpch benchmark involving three relations: \texttt{customer}, \texttt{orders}, and \texttt{lineitem}, along with a reduced form of query~\q{3}, shown in Listing~\ref{lst:tpch-q3}. The query returns an aggregate value of $2,319.5$, representing the total revenue (after discount) from the selected items. Tables~\ref{tab:lineitem}, \ref{tab:customer}, and~\ref{tab:orders} present a small subset of tuples from these relations. Tuples that contribute to the query result are highlighted; these are the focus of query explanation. Such tuples are commonly referred to as \emph{endogenous tuples} in the context of query explanation, as opposed to \emph{exogenous tuples}, which do not influence the result. Endogenous tuples are typically identified through query provenance, such as lineage.

\begin{lstlisting}[
  label={lst:tpch-q3},
  caption={Simplified \tpch \q{3}},
  frame=single,
  numbers=left,
  numbersep=5pt,
  numberstyle=\small\ttfamily,
  breaklines=true,
  breakatwhitespace=false,
  postbreak=\mbox{\textcolor{gray}{$\hookrightarrow$}\space}
]
SELECT SUM(l.extendedprice * (1 - l.discount))
AS revenue 
FROM customer c, orders o, lineitem l
WHERE c.mktsegment = 'AUTO' 
  AND c.custkey = o.custkey 
  AND l.orderkey = o.orderkey 
  AND l.shipdate > o.orderdate 
  AND o.orderkey = 23417; 
\end{lstlisting}
 
In database settings, the Shapley value provides a principled way to quantify how much each tuple contributes to a query result. It does so by averaging the tuple’s marginal contribution across all possible \emph{permutations} of the endogenous tuples. Each permutation simulates building the database one tuple at a time; the tuples that come before the target tuples define a \emph{coalition}, and the contribution of the target tuple is measured as the change in the query result when it is added. This averaging over permutations ensures a fair and context-aware attribution.

To illustrate Shapley value computation, consider the goal of computing the contribution of tuple $o_1$ from Table~\ref{tab:orders} to the final query result. Endogenous tuples in this example consist of four \texttt{lineitem} tuples ($\ell_1$ to $\ell_4$), one \texttt{customer} tuple ($c_1$), and one \texttt{orders} tuple ($o_1$). A permutation is any sequence of these tuples that represents an order in which they could be added to an initially empty database. For example:
\begin{align}
\permt = (c_1, \ell_1, \ell_2, \ell_3, o_1, \ell_4)
\end{align}

\noindent Given a permutation $\permt$ and a target tuple (e.g., $o_1$), the associated \emph{coalition} consists of $o_1$ and all tuples that appear before it in the permutation. In this case, the coalition is $\{c_1, \ell_1, \ell_2, \ell_3, o_1\}$. The marginal contribution of $o_1$ is defined as the difference in the query result when $o_1$ is included versus excluded from the coalition.

For example, the result of \q{3} on $\{c_1, \ell_1, \ell_2, \ell_3\}$ is 0, and adding $o_1$ changes the result to $1,702$. Therefore, the marginal contribution of $o_1$ in this permutation is $1,702$. The Shapley value is computed as the average of such marginal contributions across all permutations.

However, exact computation is infeasible in practice. In this small example with 6 endogenous tuples, there are exponentially many coalitions, $2^5$ as $o_1$ must be in the coalitions, and each requires executing the query twice—once with and once without the target tuple—making the process computationally intensive.\boxtheorem\end{example}

Monte Carlo sampling (\mcs) is a standard approximation technique that samples permutations or coalitions to produce an unbiased estimate without enumerating all subsets. However, its accuracy is highly sensitive to the sampling distribution: \mcs tends to oversample large coalitions and undersample small ones, even though marginal contributions often vary most at small coalition sizes. Size-stratified sampling (\ss) improves variance by dividing coalitions into strata by size and sampling uniformly within each stratum, ensuring better coverage across coalition sizes.

Yet, when applied to relational databases, both \mcs and \ss ignore \emph{schema and join structure}. Returning to Example~\ref{ex:tpch-intro}, many randomly sampled coalitions (e.g., those containing only \texttt{lineitem} tuples with no matching \texttt{customer}) cannot satisfy the join conditions and therefore yield deterministically zero or redundant marginal contributions. Moreover, because large relations such as \texttt{lineitem} dominate the tuple space, uniform sampling across tuples produces imbalanced coalitions that under-represent smaller but semantically crucial relations (e.g., \texttt{customer}). This wastes budget and inflates variance, limiting the practicality of Shapley-based attribution for relational queries.

We address these challenges with \emph{Relation-Stratified Sampling} (\rss), a schema-aware method that incorporates relational structure directly into the sampling process. Rather than stratifying coalitions solely by their size, \rss groups them by a \emph{relation-composition vector} $\mathbf{s}=(s_1,\ldots,s_r)$—the number of tuples drawn from each relation in the query. This design prioritizes strata that are more likely to satisfy join dependencies and yield informative contributions (e.g., having at least one tuple from all tables), while naturally down-weighting uninformative strata (e.g., coalitions with many \texttt{lineitem} tuples but no \texttt{customer} or \texttt{orders}). Building on this foundation, we propose an adaptive variant, \emph{Adaptive Relation-Stratified Sampling} (\arss), which reallocates sampling effort based on observed intra-stratum variance, concentrating computation where marginal contributions vary most and improving accuracy at fixed budgets.

We evaluate \rss and \arss on the \textsc{TPC-H} benchmark~\cite{tpch}, comparing against \mcs and \ss under equivalent sampling budgets. Our results show that \rss achieves consistently lower estimation error and faster error decay, and that adaptive allocation in \arss further enhances efficiency. We also implement practical optimizations, including in-database sampling (to reduce per-sample query overhead) and quantile-based binning (to limit the number of strata). Together, these techniques demonstrate that relation-aware sampling can make Shapley-based attribution feasible and accurate for realistic relational workloads. While our focus is not to deliver a production-ready system, \rss and \arss provide a principled and efficient framework for sampling-based Shapley estimation in relational settings and a foundation for future system-level advances.

Section~\ref{sec:preliminaries} introduces preliminaries and notations. Section~\ref{sec:rw} reviews related work. Section~\ref{sec:rss} presents \rss and \arss with performance optimizations. Section~\ref{sec:ex} reports experiments, and Section~\ref{sec:conclusion} concludes.

\section{Preliminaries}
\label{sec:preliminaries}

We briefly review the Shapley value and its variants in cooperative game theory, discuss sampling-based estimation methods, and describe how these concepts extend to quantify tuple contributions in databases.

\subsection{Shapley Value in Cooperative Game Theory}
\label{sec:shapley_gt}

In cooperative game theory, a game consists of a set of players \(N\) and a value function \(v: 2^N \to \mathbb{R}\) assigning a payoff to every coalition \(S \subseteq N\). The Shapley value~\cite{shapley1953value} distributes the total payoff among players based on their expected marginal contribution across all possible orders in which they could join a coalition. For a player \(i \in N\) and permutation \(\pi\) of \(N\), let \(S_i^\pi\) denote the set of players appearing before \(i\) in \(\pi\). The marginal contribution of \(i\) is 
\[
\Delta_i(S_i^\pi) = v(S_i^\pi \cup \{i\}) - v(S_i^\pi),
\]
and the Shapley value is the average of these contributions across all orderings:
\[
\phi_i(v) = \frac{1}{|N|!} \sum_{\pi \in \Pi(N)} \Delta_i(S_i^\pi).
\]
Equivalently, the Shapley value can be written as a sum over all coalitions weighted by their number of corresponding permutations:
\[
\phi_i(v) = \sum_{S \subseteq N \setminus \{i\}} 
w(S)\,\Delta_i(S).
\]

with the weight%
\begin{align}
w(S)=\frac{|S|! (|N|-|S|-1)!}{|N|!}. \label{eq:w}  
\end{align}

\noindent This formulation satisfies key axioms—efficiency, symmetry, additivity, and the null-player property—making the Shapley value the unique fair division rule in cooperative games.

It is important to note that the Shapley value is not limited to positive numbers. In general, the characteristic function \(v(S)\) may take any real value, including negative ones. If adding a player decreases the value of a coalition, then the marginal contribution \(v(S \cup \{i\}) - v(S)\) becomes negative, and the player's Shapley value can also be negative. This is fully valid within the Shapley framework.

\subsection{Sampling-Based Estimation}
\label{sec:sampling}

Exact Shapley computation is infeasible for large \(N\) due to the exponential number of coalitions. Monte Carlo sampling (\mcs)~\cite{castro2009polynomial} offers a simple, unbiased approximation by averaging marginal contributions over random coalitions or permutations. Given \(m\) samples, the estimate is
\begin{align}\hat{\phi}_i(v) = \frac{2^{|N|-1}}{m} \sum_{j=1}^{m} w(S_j)\, \Delta_i(S_j),\label{fm:ss-agg}
\end{align}
where \(w(S_j)\) is the Shapley weight in Eq~\ref{eq:w}. The variance of this estimator decreases as \(O(1/\sqrt{m})\), but \mcs tends to oversample large coalitions, causing high variance when small coalitions dominate contributions.

Stratified sampling (\ss)~\cite{maleki2014} reduces variance by grouping coalitions into strata based on size and allocating samples within each group. The total sample budget is distributed either proportionally to stratum size or according to variance-aware rules such as Neyman allocation~\cite{zhang2023efficient}. When stratum variances are unknown, adaptive schemes~\cite{castro2017improving,burgess2021approximating} estimate them during sampling and adjust allocation dynamically. These approaches achieve lower variance without increasing sample size, forming the basis for our relation-aware adaptive sampling strategy.

\subsection{Shapley-Like Values}
\label{sec:banzhaf}

Several generalizations of the Shapley value modify its weighting to trade off fairness and computational cost. The Banzhaf value~\cite{banzhaf1965weighted} assigns equal weight to all coalitions,
\[
\beta_i(v) = \frac{1}{2^{|N|-1}} \sum_{S \subseteq N \setminus \{i\}} \Delta_i(S),
\]
simplifying computation while relaxing the efficiency axiom. The Owen value~\cite{owen1977values} extends Shapley to hierarchical settings where players belong to groups, first distributing value among groups, then among members. The Beta-Shapley value~\cite{kwon2022beta} introduces a Beta distribution over coalition sizes, allowing emphasis on smaller or larger coalitions. Sampling-based estimation naturally supports these variants by adjusting the weighting or sampling distribution, making it a flexible framework for computing different attribution models.

\subsection{Shapley Value for Database Queries}
\label{sec:db-preliminaries}

We adapt the Shapley framework to measure how database tuples contribute to query results~\cite{livshits2021shapley}. Let \(D\) denote a database instance (the set of tuples) and \(Q\) a query producing a scalar output, such as an aggregate value or Boolean result. The value function is \(v(S) = Q(S)\), representing the query output on any subset \(S \subseteq D\). For a tuple \(t \in D\), its marginal contribution to coalition \(S\) is
\[
\Delta_t(S) = Q(S) - Q(S \setminus \{t\}),
\]
and its Shapley value is
\[
\phi_t(Q, D) = \sum_{S \subseteq D \setminus \{t\}} 
\frac{|S|! (|D|-|S|-1)!}{|D|!} \, \Delta_t(S).
\]
This quantifies how much \(t\) contributes to the query outcome on average across all possible subsets of other tuples. The same formulation extends to alternative attribution models (e.g., Banzhaf or Beta-Shapley) by modifying the coalition weighting. Sampling-based estimators, such as those discussed above, provide a tractable means of approximating these values through repeated query evaluation, making Shapley analysis feasible for realistic databases.

\section{Related Work}
\label{sec:rw}

Research on Shapley-like values spans two major directions: sampling-based estimation methods and their applications to databases. The first line of work focuses on scalable approximation techniques for computing Shapley values, while the second investigates their use in explaining, valuing, and repairing data through query-level attributions.

Computing the Shapley value exactly is intractable because it requires evaluating all possible coalitions of players, an exponential number. Early work introduced Monte Carlo sampling (\mcs) as a practical estimator~\cite{mann1960values}, later formalized with asymptotic guarantees by~\cite{castro2009polynomial}. Stratified sampling (\ss) was proposed to mitigate the high variance of \mcs by grouping coalitions based on size and allocating samples across strata~\cite{maleki2014}. More recent methods refine allocation strategies to account for intra-stratum variability. The classical Neyman allocation rule minimizes variance by assigning more samples to high-variance strata~\cite{zhang2023efficient}, while adaptive two-stage schemes~\cite{castro2017improving} and empirical Bernstein methods~\cite{burgess2021approximating} dynamically adjust allocation during sampling. These approaches improve efficiency when stratum variances differ sharply, a property that naturally arises in relational data where certain coalition structures dominate contribution variance.

Beyond standard random sampling, kernel-based and quasi–Monte Carlo (QMC) extensions~\cite{mitchell2022sampling} introduce geometric structure into the sampling process. By formulating Shapley estimation as an integration problem in a reproducing kernel Hilbert space, they adaptively select low-discrepancy permutations to accelerate convergence. While effective in structured domains, these methods are less practical in databases, where each permutation entails repeated query execution. Our work instead focuses on relation-aware stratification, which captures structural dependencies among tuples without sequential evaluation, providing a more tractable alternative for relational settings.

The use of Shapley values in databases is a more recent development. Livshits et al.~\cite{livshits2022shapley} and Deutch et al.~\cite{deutch2021explanations} applied Shapley-based reasoning to data cleaning and explanation, showing how tuple contributions can quantify repair influence or constraint responsibility. Tian et al.~\cite{tian2022data} and Si and Pei~\cite{si2024counterfactual} used Shapley attribution for fair data pricing and counterfactual valuation. Davidson et al.~\cite{davidson2022shapgraph} introduced \textit{ShapGraph}, integrating Shapley values with provenance graphs for query explanation, while Bienvenu et al.~\cite{bienvenu2024shapley} extended Shapley reasoning to ontology-mediated queries. Karmakar et al.~\cite{karmakar2024expected} and Kara et al.~\cite{kara2024shapley} explored expected Shapley computation in probabilistic databases and its connection to model counting, respectively.

Recent computational studies aim to make these analyses tractable. Livshits et al.~\cite{livshits2021shapley} characterized the complexity of Shapley computation for conjunctive and aggregate queries, showing polynomial-time results only for hierarchical queries. Deutch et al.~\cite{deutch2022computing} reduced Shapley computation to probabilistic query evaluation via knowledge compilation, while Abramovich et al.~\cite{abramovich2024banzhaf,abramovich2025advancing} unified Shapley and Banzhaf-based influence scores using provenance factorization and approximation schemes. Despite their progress, these approaches depend on lineage extraction or specialized query classes, limiting scalability.

Our work differs in that it treats the database as a black box and focuses on sampling-based Shapley estimation that directly leverages relational structure. By introducing relation-aware stratification and adaptive allocation, we extend classical sampling methods into a practical framework for scalable, structure-conscious Shapley value estimation in relational databases.

\section{Adaptive Relation-Stratified Sampling for Shapley Estimation of Database Queries}
\label{sec:rss-main}

This section presents \emph{Relation-Stratified Sampling} (\rss) and its adaptive variant (\arss) for estimating Shapley values over database queries. We first define the stratification scheme based on relation-level vectors and explain how it leverages schema structure to reduce variance and avoid uninformative coalitions. We then describe an adaptive allocation strategy that reallocates samples across strata based on observed variability, followed by implementation details that enable practical execution inside a relational system. We conclude with a complexity analysis and system-level optimizations that reduce redundant work and lower per-sample query cost.

\subsection{\rss: Relation-Stratified Sampling}
\label{sec:rss}

Classical Monte Carlo sampling and size-based stratified sampling become inefficient in relational settings because they ignore schema structure and join requirements. Many sampled coalitions fail to contain the relational components needed for the query to produce a nonzero contribution; these samples waste budget and inflate variance. \rss addresses this by grouping coalitions not by their total size but by their \emph{relational composition}—the number of tuples drawn from each participating relation. This stratification captures the structural constraints induced by joins and selections and concentrates sampling on coalitions that are more likely to be informative.

Let $(D,Q,t)$ be an explanation setting as in Section~\ref{sec:db-preliminaries}. Assume that \(Q\) involves \(r\) endogenous relations \(R_1,\ldots,R_r\) and denote by \(\Endg{Q}{i}\subseteq R_i\) the set of endogenous tuples from relation \(R_i\). We assume \(t\in \EndgR{Q}=\bigcup_{i=1}^r \Endg{Q}{i}\); otherwise \(\phi_t(Q,D)=0\). A coalition is summarized by a \emph{relation vector} \(\mathbf{s}=(s_1,\ldots,s_r)\), where \(s_i\in\{0,\ldots,|\Endg{Q}{i}|\}\) counts how many tuples are drawn from \(\Endg{Q}{i}\). Each distinct \(\mathbf{s}\) defines a stratum, and the stratification space is
\[
\mathcal{G}=\{(s_1,\ldots,s_r)\mid 0\le s_i\le |\Endg{Q}{i}|\}.
\]
Within a stratum \(\mathbf{s}\), all coalitions contain exactly \(s_i\) tuples from relation \(R_i\) for every \(i\). The number of coalitions in \(\mathbf{s}\) is \(\prod_{i=1}^r \binom{|\Endg{Q}{i}|}{s_i}\). This counting directly exposes the combinatorial skew that arises from large relations: strata that over-select from a single large relation without covering the others are numerous but usually uninformative for join-heavy queries. \rss exploits this observation by sampling uniformly inside each \(\mathbf{s}\) but allocating fewer samples to relation vectors that are structurally unlikely to join, and more to balanced vectors that cover the join graph. In its simplest form, \rss assigns a proportional share of the budget to each stratum according to its size,
\[
w_{\mathbf{s}}=\frac{\prod_{i=1}^r \binom{|\Endg{Q}{i}|}{s_i}}{2^{|N|}-1},\quad N=\bigcup_{i=1}^r \Endg{Q}{i},
\]
and estimates the Shapley value using the stratified estimator in Eq.~\eqref{fm:ss-agg}. In practice, proportionality serves as a neutral prior when no variance information is available; the full benefit of relation-aware stratification emerges once we adapt allocations to the observed variability of marginal contributions in each \(\mathbf{s}\).

\subsection{\arss: Adaptive Allocation over Relation Vectors}
\label{sec:rss-allocation}

Variance can differ sharply across relation vectors. Strata that cover key join paths or selective predicates often exhibit large and heterogeneous marginal contributions, while strata missing critical relations tend to yield zeros with low variance. Allocating samples proportionally to stratum size ignores this heterogeneity and expends substantial budget on low-impact strata. \arss addresses this by updating the allocation over time using empirical variance estimates gathered during sampling.

We divide the total budget \(m\) into \(k\) cycles of size \(m/k\). The first cycle uses the proportional rule as a cold start, drawing a small number of samples from every \(\mathbf{s}\in\mathcal{G}\) and recording the observed marginal contributions for \(t\). After the first cycle, we compute standard deviations \(\hat{\sigma}_{\mathbf{s}}\) for each stratum from the collected samples. Subsequent cycles allocate their budget via a Neyman-style rule that increases the share of samples for strata with larger \(\hat{\sigma}_{\mathbf{s}}\), while maintaining a small exploration floor so no stratum is starved. After each cycle, we update \(\hat{\sigma}_{\mathbf{s}}\) and reapply the rule. This batchwise procedure balances stability and responsiveness: larger batches lower allocation overhead, while smaller batches react more quickly to the evolving variance profile. Our experiments show that moderate values of \(k\) deliver faster error decay than static allocations without incurring significant overhead.

Relation-aware stratification and adaptive allocation reinforce each other. The relational grouping creates strata that are more homogeneous with respect to their structural ability to contribute, lowering intra-stratum variance; adaptive allocation then focuses effort within \(\mathcal{G}\) precisely where contributions remain most variable. The result is an estimator that preserves unbiasedness while achieving lower variance and faster convergence than \mcs and size-only \ss under the same budget.

\subsection{Implementation Details}

We implement \rss/\arss as a black-box procedure around the database. Endogenous tuples are identified once using a provenance tool (e.g., Perm~\cite{glavic2009perm}) to obtain \(\EndgR{Q}\) and its partition by relation. Each tuple carries a globally unique identifier, enabling us to materialize sampled coalitions through simple filters. For a chosen relation vector \(\mathbf{s}\), we uniformly select \(s_i\) identifiers from \(\Endg{Q}{i}\) and evaluate two queries: one that includes the target tuple \(t\) and one that excludes it. The difference of the two results is the marginal contribution for that sample. We repeat this process according to the current allocation over \(\mathcal{G}\), accumulating per-stratum statistics for \arss and aggregating all samples into the final stratified estimate.

A straightforward implementation rewrites the original SQL with \texttt{IN} predicates over the sampled identifiers. While simple, repeated rewriting and optimization can become expensive for large coalitions or complex queries. We therefore pair the sampler with view-based compilation to precompute shared subexpressions and joins once and evaluate much lighter queries per sample; Section~\ref{sec:opt} details these optimizations. We also cache previously evaluated coalitions to avoid redundant work and apply static pruning rules to skip strata or coalitions that cannot affect the result (for example, identifier sets that cannot join to the target).

\subsection{Analysis}
\label{sec:rss-analysis}

The runtime of \arss is dominated by query evaluation. Let \(q\) denote the average cost of one query execution for a sampled coalition. Each sample requires two query evaluations (with and without \(t\)), but this factor is constant and absorbed into \(q\). With total budget \(m\) spread over \(k\) cycles, the sampling cost is \(O(m\,q)\). Allocation and variance updates require iterating over all strata in \(\mathcal{G}\) once per cycle, contributing \(O(k\,|\mathcal{G}|)\). Since \(m\) is typically much larger than \(k\), and because query evaluation is orders of magnitude costlier than updating a few per-stratum statistics, the overall complexity is governed by \(O(m\,q)\). The chief variance benefit comes from two sources: relation-aware grouping reduces intra-stratum variability by aligning samples with the query’s join structure, and adaptive allocation concentrates budget in the few relation vectors that continue to exhibit high variance. Empirically, this produces lower error and faster error decay than \mcs and size-only \ss for the same \(m\).

The main challenge is that \(|\mathcal{G}|\) can grow with the number of relations and endogenous tuples per relation. In practice, many relation vectors are uninformative because they cannot satisfy the join graph or they omit selective relations entirely. We therefore collapse or remove such vectors using static analysis of the query and observed join selectivity, and we optionally bin \(s_i\) values into quantiles per relation to coarsen \(\mathcal{G}\) without sacrificing structural fidelity. These reductions decrease allocation overhead and improve the stability of variance estimates in \arss.

\section{Experiments}
\label{sec:ex}

The objective of our experimental study is to assess whether relation-aware stratification and adaptive allocation provide tangible benefits for Shapley estimation on relational queries. We compare \rss and its adaptive variant \arss against classical size-based \ss and its adaptive counterpart \ass. The evaluation includes an ablation that isolates the effect of relation-based stratification from adaptive allocation, thereby clarifying where the gains arise. We use real \tpch workloads to ensure practical relevance. Our goal is not to present a fully scalable production system for exact Shapley computation; rather, we demonstrate that relation-aware, variance-driven sampling substantially improves estimation accuracy and convergence under realistic query costs. We conclude with observations on limitations and directions for future work.

\subsection{Experimental Setup}
\label{sec:setup}

All methods were implemented in Java 22.0.2 (Eclipse Temurin) and interfaced with PostgreSQL through JDBC. Experiments ran on an 11th Gen Intel\textsuperscript{\textregistered} Core\textsuperscript{TM} i7-1185G7 (8 logical threads) with 32\,GB DDR4. Tuple-level provenance was obtained using \perm~\cite{glavic2009perm}. We evaluate on \tpch, a standard decision-support benchmark with multi-relation joins and aggregates. We use the 1\,GB scale to focus on estimator behavior rather than distributed scalability. Table~\ref{tab:tpch-schema} summarizes relation sizes and attributes.

\begin{table}[t!]
    \centering
    \begin{tabular}{lrr}
            \toprule
            \texttt{relation} & \nrel & \ncolmn \\
            \midrule
            \texttt{lineitem} & 6{,}001{,}215 & 16 \\
            \texttt{orders} & 1{,}500{,}000 & 9 \\
            \texttt{partsupp} & 800{,}000 & 5 \\
            \texttt{part} & 200{,}000 & 9 \\
            \texttt{customer} & 150{,}000 & 8 \\
            \texttt{supplier} & 10{,}000 & 7 \\
            \texttt{nation} & 25 & 4 \\
            \texttt{region} & 5 & 3 \\
            \bottomrule
        \end{tabular}
        \caption{\tpch 1\,GB dataset overview.}
        \label{tab:tpch-schema}
\end{table}

We add a unique \texttt{id} attribute to each base relation to enable sampling and filtered evaluation. We consider a subset of \tpch queries with moderate provenance sizes to allow repeated estimation and controlled comparisons. Table~\ref{tab:tpch-query-info} reports query-level statistics and \perm costs; highlighted queries are used in our study. Full schema, queries, and scripts are in our repository~\cite{alizad2025gametheoretic}.

\begin{table}[t!]
  \centering
    \begingroup
      \setlength{\tabcolsep}{4pt}        
      \renewcommand{\arraystretch}{1.0}  
      \begin{tabular}{@{}l r r l r@{}}   
        \toprule
        & \nrel & \nans & \psize \texttt{range} & \ptime \\
        \midrule
        \q{1}  & 1 & 4        & [38{,}854, 3{,}004{,}998] & \formattime{21.35}{s}{2} \\
        \rowcolor{lightred}\q{2}  & 5 & 460      & [5, 11]                  & \formattime{0.36}{s}{2} \\
        \rowcolor{lightred}\q{3}  & 3 & 11{,}403 & [3, 9]                   & \formattime{3.46}{s}{2} \\
        \q{4}  & 2 & 5        & [38{,}663, 39{,}161]      & \formattime{5.19}{s}{2} \\
        \q{5}  & 6 & 5        & [4{,}155, 4{,}530]        & \formattime{2.08}{s}{2} \\
        \q{6}  & 1 & 1        & 44{,}983                  & \formattime{4.36}{s}{2} \\
        \q{7}  & 5 & 2        & [4{,}261, 4{,}299]        & \formattime{3.97}{s}{2} \\
        \q{8}  & 7 & 2        & [109{,}570, 111{,}064]    & \formattime{5.15}{s}{2} \\
        \q{9}  & 6 & 175      & [-, -]                    & - \\
        \rowcolor{lightred}\q{10} & 4 & 20       & [14, 27]                 & \formattime{5.57}{s}{2} \\
        \q{11} & 3 & 1{,}056  & [31{,}834, 31{,}835]      & \formattime{0.73}{s}{2} \\
        \q{12} & 2 & 1        & 31{,}107                  & \formattime{4.24}{s}{2} \\
        \q{13} & 2 & 42       & [-, -]                    & - \\
        \q{14} & 2 & 1        & 141{,}477                 & \formattime{2.96}{s}{2} \\
        \q{15} & 2 & 1        & [-, -]                    & - \\
        \q{16} & 3 & 1{,}339  & [-, -]                    & - \\
        \q{17} & 2 & 1        & [-, -]                    & - \\
        \rowcolor{lightred}\q{18} & 3 & 57       & [9, 9]                   & \formattime{81.21}{s}{2} \\
        \q{19} & 2 & 1        & 188                       & \formattime{4.49}{s}{2} \\
        \q{20} & 3 & 108{,}400 & [-, -]                   & - \\
        \rowcolor{lightred}\q{21} & 4 & 393      & [13, 79]                 & \formattime{61.6}{s}{2} \\
        \q{22} & 3 & 7        & [-, -]                    & - \\
        \bottomrule
      \end{tabular}
    \endgroup
    \caption{Query statistics for \tpch. Highlighted queries are used in experiments.}
    \label{tab:tpch-query-info}
\end{table}

\subsection{Measures and Baselines}
\label{sec:measures}

Accuracy is measured by mean relative error (\mre). Given the exact value \(v^*\) and estimates \(\hat{v}_1,\ldots,\hat{v}_k\) from repeated runs with a fixed budget, we compute
\[
\mre = \frac{1}{k} \sum_{i=1}^{k} \frac{|\hat{v}_i - v^*|}{|v^*|}.
\]
This normalization allows comparison across queries and provenance sizes. We also report end-to-end runtime (\ttime) and in-database query time (\qtime) to separate orchestration overhead from evaluation cost. We compare \rss and \arss against size-based \ss and \ass. Plain \mcs is omitted due to inferior accuracy and cost on multi-join workloads relative to stratified methods.

\subsection{Scenarios and Settings}

We evaluate two scenarios. The first reports \mre for representative settings where exact values can be computed quickly; these verify estimator correctness and illustrate low-variance strata induced by relation-aware grouping. The second examines convergence as the sample budget grows, emphasizing settings where exact enumeration is expensive and approximation quality matters most.

Selected settings come from \q{2}, \q{3}, \q{10}, \q{18}, and \q{21}. On average, each query yields \(\approx\)2.4k answers with \(\approx\)20 provenance tuples per answer, implying hundreds of thousands of potential estimation tasks. We therefore sample representative cases. Low-cost settings (Table~\ref{tab:low-cost-settings}) have small provenance and allow exact computation in seconds to minutes. High-cost settings (Table~\ref{tab:settings-table}) involve larger or more complex provenance; exact enumeration can take hours, reflecting realistic analytics. In both regimes, we keep the environment identical and vary only the estimator and sample budget.

\begin{table}[h!]
\centering
\small
\renewcommand{\arraystretch}{1.0}
\begingroup
  \setlength{\tabcolsep}{3.5pt} 
  \begin{tabular}{@{}l@{\hspace{6pt}}l@{\hspace{8pt}}l@{\hspace{8pt}}r@{\hspace{6pt}}r@{}}
  \toprule
  \textbf{Setting} & \textbf{Query} & \textbf{Size Vector} & \textbf{Shapley Value} & \textbf{Shapley Time} \\
  \midrule
  \setting{2}{L1}  & \q{2}  & (3, 1, \underline{3}, 1, 3) & 0.0409     & \timeinseconds{295} s \\
  \setting{3}{L1}  & \q{3}  & (1, 7, \underline{1})        & 81462.9280 & \timeinseconds{119} s \\
  \setting{10}{L1} & \q{10} & (1, 9, 1, \underline{3})     & 8798.1953  & \timeinseconds{3655} s \\
  \setting{18}{L1} & \q{18} & (1, \underline{7}, 1)        & 16.6448    & \timeinseconds{178} s \\
  \bottomrule
  \end{tabular}
\endgroup
\caption{Low-cost settings used for verified comparisons.}
\label{tab:low-cost-settings}
\end{table}

\begin{table}[h!]
\centering
\small
\renewcommand{\arraystretch}{1.0}
\begingroup
  \setlength{\tabcolsep}{3.5pt} 
  \begin{tabular}{@{}l@{\hspace{6pt}}l@{\hspace{8pt}}l@{\hspace{8pt}}r@{\hspace{6pt}}r@{}}
  \toprule
  \textbf{Setting} & \textbf{Query} & \textbf{Size Vector} & \textbf{Shapley Value} & \textbf{Shapley Time} \\
  \midrule
  \setting{10}{H1} & \q{10} & (1, \underline{20}, 1, 5)           & 61657.9447 & $\approx$6 hr \\
  \setting{10}{H2} & \q{10} & (1, \underline{17}, 1, 5)           & 321.3894   & $\approx$50 min \\
  \setting{21}{H1} & \q{21} & (12, 1, \underline{5}, 1)           & 0.2333     & $\approx$2.5 min \\
  \setting{21}{H2} & \q{21} & (16, 1, 7, \underline{1})           & 1.4666     & $\approx$5 hr \\
  \setting{21}{H3} & \q{21} & (\underline{13}, 1, 6, \underline{1}) & 0.1999   & $\approx$9 min \\
  \bottomrule
  \end{tabular}
\endgroup
\caption{High-cost settings representative of realistic workloads.}
\label{tab:settings-table}
\end{table}

\subsection{Results}
\label{sec:exp-res}

We present results in two parts: accuracy at fixed budgets and convergence behavior as samples increase. Tables~\ref{tab:q2-1} and~\ref{tab:q2-2} summarize accuracy and time for low- and high-cost settings, respectively, across 1k, 10k, and 100k samples.

On low-cost settings (Table~\ref{tab:q2-1}), \arss delivers the lowest \mre in nearly all cases, often matching exact values when relation-aware strata collapse variance. For \setting{10}{L1} at 100k samples, \arss attains \(\mre=0.002\), outpacing \rss (0.003), \ss (0.006), and \ass (0.072). In \setting{3}{L1} and \setting{18}{L1}, both \rss and \arss converge to exactness because relation vectors align tightly with the query’s join structure, yielding near-zero intra-stratum variance. When variance persists, \arss maintains a clear advantage by reallocating samples to the most informative relation vectors. Runtime is also competitive: because query evaluation dominates cost, reducing wasted samples in uninformative strata translates directly to lower \qtime and, consequently, lower \ttime.

\begin{table*}[h!]
\centering
\small
\setlength{\tabcolsep}{2pt}
\renewcommand{\arraystretch}{0.9}
\begin{tabular}{ll c c c c c c c c c c c c }
\toprule
\multirow{2}{*}{\textbf{Set.}} & \multirow{2}{*}{\textbf{Meth.}}
& \multicolumn{3}{c}{\textbf{Estimation}}
& \multicolumn{3}{c}{\textbf{\mre}}
& \multicolumn{3}{c}{\textbf{\ttime (s)}}
& \multicolumn{3}{c}{\textbf{\qtime (s)}} \\
\cmidrule(lr){3-5} \cmidrule(lr){6-8} \cmidrule(lr){9-11} \cmidrule(lr){12-14}
& & 1k & 10k & 100k
  & 1k & 10k & 100k
  & 1k & 10k & 100k
  & 1k & 10k & 100k \\
\midrule
\multirow{4}{*}{\setting{2}{L1}}
    & \ss   & 0.0467 & 0.0425 & 0.0404 & 0.171 & 0.040 & 0.012 & \timeinseconds{255} & \timeinseconds{2313} & \timeinseconds{34422} & \timeinseconds{232} & \timeinseconds{2227} & \timeinseconds{33191} \\
    & \ass  & 0.0471 & 0.0414 & 0.0410 & 0.177 & \shl{0.018} & 0.011 & \shl{\timeinseconds{226}} & \shl{\timeinseconds{641}} & \hl{\timeinseconds{6735}} & \shl{\timeinseconds{192}} & \shl{\timeinseconds{592}} & \hl{\timeinseconds{6313}} \\
    & \rss  & 0.0366 & 0.0401 & 0.0407 & \shl{0.141} & 0.026 & \shl{0.008} & \timeinseconds{284} & \timeinseconds{2385} & \timeinseconds{34487} & \timeinseconds{250} & \timeinseconds{2288} & \timeinseconds{33478} \\
    & \arss & 0.0394 & 0.0405 & 0.0409 & \hl{0.090} & \hl{0.013} & \hl{0.005} & \hl{\timeinseconds{176}} & \hl{\timeinseconds{669}} & \shl{\timeinseconds{7729}} & \hl{\timeinseconds{105}} & \hl{\timeinseconds{567}} & \shl{\timeinseconds{6783}} \\
\midrule
\multirow{4}{*}{\setting{3}{L1}}
    & \ss   & 1851.62 & 1812.98 & 1822.10 & 0.023 & 0.009 & 0.003 & \timeinseconds{312} & \timeinseconds{3645} & \timeinseconds{14115} & \timeinseconds{253} & \timeinseconds{3500} & \timeinseconds{13285} \\
    & \ass  & 1675.81 & 1825.38 & 1824.42 & 0.081 & \shl{0.006} & 0.004 & \hl{\timeinseconds{61}} & \shl{\timeinseconds{1950}} & \shl{\timeinseconds{2745}} & \hl{\timeinseconds{41}} & \shl{\timeinseconds{271}} & \shl{\timeinseconds{2500}} \\
    & \rss  & 1822.66 & 1822.66 & 1822.66 & 0.000 & 0.000 & 0.000 & \timeinseconds{303} & \timeinseconds{2032} & \timeinseconds{12297} & \timeinseconds{266} & \timeinseconds{1674} & \timeinseconds{11733} \\
    & \arss & 1822.66 & 1822.66 & 1822.66 & \hl{0.000} & \hl{0.000} & \hl{0.000} & \shl{\timeinseconds{82}} & \hl{\timeinseconds{278}} & \hl{\timeinseconds{2649}} & \shl{\timeinseconds{50}} & \hl{\timeinseconds{237}} & \hl{\timeinseconds{2318}} \\
\midrule
\multirow{4}{*}{\setting{10}{L1}}
    & \ss   & 5311.38 & 7803.13 & 8840.18 & 0.396 & 0.120 & 0.006 & \shl{\timeinseconds{381}} & \timeinseconds{3693} & \timeinseconds{28038} & \shl{\timeinseconds{333}} & \timeinseconds{3473} & \timeinseconds{26828} \\
    & \ass  & 4001.13 & 7242.23 & 8199.56 & 0.545 & 0.177 & 0.072 & \timeinseconds{592} & \hl{\timeinseconds{1810}} & \hl{\timeinseconds{6636}} & \timeinseconds{532} & \hl{\timeinseconds{1681}} & \hl{\timeinseconds{5914}} \\
    & \rss  & 4977.56 & 8236.83 & 8782.43 & 0.434 & \shl{0.069} & \shl{0.003} & \hl{\timeinseconds{348}} & \timeinseconds{3445} & \timeinseconds{32691} & \hl{\timeinseconds{297}} & \timeinseconds{3244} & \timeinseconds{31195} \\
    & \arss & 8903.07 & 8793.92 & 8811.11 & \hl{0.035} & \hl{0.022} & \hl{0.002} & \timeinseconds{563} & \shl{\timeinseconds{2755}} & \shl{\timeinseconds{7116}} & \timeinseconds{423} & \shl{\timeinseconds{2514}} & \shl{\timeinseconds{6319}} \\
\midrule
\multirow{4}{*}{\setting{18}{L1}}
    & \ss   & 16.66 & 16.75 & 16.60 & 0.022 & 0.009 & 0.004 & \timeinseconds{415} & \timeinseconds{3052} & \timeinseconds{23391} & \timeinseconds{332} & \timeinseconds{2772} & \timeinseconds{22231} \\
    & \ass  & 14.01 & 16.68 & 16.66 & 0.158 & \shl{0.009} & \shl{0.003} & \hl{\timeinseconds{129}} & \hl{\timeinseconds{581}} & \hl{\timeinseconds{4274}} & \hl{\timeinseconds{99}} & \hl{\timeinseconds{528}} & \hl{\timeinseconds{3961}} \\
    & \rss  & 16.65 & 16.64 & 16.64 & 0.000 & 0.000 & 0.000 & \timeinseconds{288} & \timeinseconds{2899} & \timeinseconds{21907} & \timeinseconds{252} & \timeinseconds{2731} & \timeinseconds{21108} \\
    & \arss & 16.65 & 16.64 & 16.64 & \hl{0.000} & \hl{0.000} & \hl{0.000} & \shl{\timeinseconds{176}} & \shl{\timeinseconds{632}} & \shl{\timeinseconds{5145}} & \shl{\timeinseconds{132}} & \shl{\timeinseconds{537}} & \shl{\timeinseconds{4732}} \\
\bottomrule
\end{tabular}
\caption{Low-cost settings: accuracy and runtime across sample budgets. Best values highlighted.}
\label{tab:q2-1}
\end{table*}

High-cost settings (Table~\ref{tab:q2-2}) amplify these trends. \arss achieves the lowest \mre across all budgets, often by large margins. In \setting{10}{H1} and \setting{10}{H2}, \arss yields \(\mre\in[0.006,0.027]\) at 1k samples and improves further with more samples, while the baselines remain one to two orders of magnitude worse even at 100k. Size-only \rss can underperform when the number of relation vectors is large and uniform allocation starves informative strata; \arss avoids this by shifting budget according to observed variance. Although \arss introduces modest allocation overhead and maintains more per-stratum state, its query time remains competitive because it avoids spending budget on uninformative coalitions.

\begin{table*}[h!]
\centering
\small
\setlength{\tabcolsep}{2pt}
\renewcommand{\arraystretch}{0.9}
\begin{tabular}{ll c c c c c c c c c c c c }
\toprule
\multirow{2}{*}{\textbf{Set.}} & \multirow{2}{*}{\textbf{Meth.}}
& \multicolumn{3}{c}{\textbf{Estimation}}
& \multicolumn{3}{c}{\textbf{\mre}}
& \multicolumn{3}{c}{\textbf{\ttime (s)}}
& \multicolumn{3}{c}{\textbf{\qtime (s)}} \\
\cmidrule(lr){3-5} \cmidrule(lr){6-8} \cmidrule(lr){9-11} \cmidrule(lr){12-14}
& & 1k & 10k & 100k
  & 1k & 10k & 100k
  & 1k & 10k & 100k
  & 1k & 10k & 100k \\
\midrule
\multirow{4}{*}{\setting{10}{H1}}
    & \ss   & 27051.83 & 31819.80 & 34234.93 & 0.561 & 0.484 & 0.445 & \timeinseconds{908} & \shl{\timeinseconds{4520}} & \hl{\timeinseconds{34002}} & \timeinseconds{665} & \shl{\timeinseconds{3812}} & \hl{\timeinseconds{31867}} \\
    & \ass  & 16874.95 & 25075.85 & 32020.03 & 0.726 & 0.593 & 0.481 & \hl{\timeinseconds{474}} & \hl{\timeinseconds{3498}} & \timeinseconds{41201} & \hl{\timeinseconds{402}} & \hl{\timeinseconds{3272}} & \timeinseconds{39191} \\
    & \rss  & 18054.63 & 32123.42 & 36143.81 & 0.707 & 0.479 & 0.414 & \shl{\timeinseconds{569}} & \timeinseconds{5204} & \shl{\timeinseconds{37156}} & \shl{\timeinseconds{436}} & \timeinseconds{4620} & \shl{\timeinseconds{34118}} \\
    & \arss & 62362.60 & 61374.83 & 61505.00 & \hl{0.011} & \hl{0.010} & \hl{0.006} & \timeinseconds{1474} & \timeinseconds{7904} & \timeinseconds{71493} & \timeinseconds{952} & \timeinseconds{7098} & \timeinseconds{67283} \\
\midrule
\multirow{4}{*}{\setting{10}{H2}}
    & \ss   & 133.61 & 184.16 & 251.88 & 0.584 & 0.427 & 0.216 & \hl{\timeinseconds{360}} & \hl{\timeinseconds{3836}} & \hl{\timeinseconds{23462}} & \hl{\timeinseconds{307}} & \hl{\timeinseconds{3552}} & \hl{\timeinseconds{22264}} \\
    & \ass  & 99.85  & 156.78 & 199.60 & 0.689 & 0.512 & 0.379 & \timeinseconds{613} & \shl{\timeinseconds{4121}} & \shl{\timeinseconds{26173}} & \timeinseconds{538} & \shl{\timeinseconds{3824}} & \shl{\timeinseconds{24601}} \\
    & \rss  & 105.14 & 162.91 & 213.42 & 0.673 & 0.493 & 0.336 & \shl{\timeinseconds{530}} & \timeinseconds{5701} & \timeinseconds{28631} & \shl{\timeinseconds{450}} & \timeinseconds{5286} & \timeinseconds{26373} \\
    & \arss & 323.08 & 320.41 & 320.30 & \hl{0.027} & \hl{0.019} & \hl{0.013} & \timeinseconds{1210} & \timeinseconds{6329} & \timeinseconds{45829} & \timeinseconds{694} & \timeinseconds{5594} & \timeinseconds{42912} \\
\midrule
\multirow{4}{*}{\setting{21}{H1}}
    & \ss   & 0.1103 & 0.1593 & 0.1878 & 0.527 & 0.317 & 0.195 & \hl{\timeinseconds{538}} & \timeinseconds{7003} & \timeinseconds{55395} & \hl{\timeinseconds{485}} & \timeinseconds{6468} & \timeinseconds{53290} \\
    & \ass  & 0.0808 & 0.1224 & 0.1588 & 0.653 & 0.475 & 0.319 & \timeinseconds{660} & \hl{\timeinseconds{5779}} & \shl{\timeinseconds{53596}} & \timeinseconds{600} & \hl{\timeinseconds{5573}} & \shl{\timeinseconds{52088}} \\
    & \rss  & 0.0867 & 0.1587 & 0.1888 & 0.628 & 0.320 & 0.191 & \shl{\timeinseconds{594}} & \shl{\timeinseconds{5785}} & \hl{\timeinseconds{50550}} & \shl{\timeinseconds{530}} & \hl{\timeinseconds{5170}} & \hl{\timeinseconds{48648}} \\
    & \arss & 0.2359 & 0.2325 & 0.2372 & \hl{0.045} & \hl{0.015} & \hl{0.012} & \timeinseconds{721} & \timeinseconds{8688} & \timeinseconds{83019} & \timeinseconds{552} & \timeinseconds{8247} & \timeinseconds{80617} \\
\midrule
\multirow{4}{*}{\setting{21}{H2}}
    & \ss   & 0.3602 & 0.7436 & 0.8769 & 0.754 & 0.493 & 0.402 & \hl{\timeinseconds{912}} & \hl{\timeinseconds{11167}} & \timeinseconds{113964} & \hl{\timeinseconds{845}} & \hl{\timeinseconds{10701}} & \timeinseconds{111317} \\
    & \ass  & 0.8380 & 0.5040 & 0.7061 & 0.646 & 0.656 & 0.519 & \timeinseconds{1157} & \shl{\timeinseconds{11786}} & \shl{\timeinseconds{112419}} & \timeinseconds{969} & \shl{\timeinseconds{11479}} & \shl{\timeinseconds{110178}} \\
    & \rss  & 0.5349 & 0.5771 & 0.8816 & 0.635 & 0.607 & 0.399 & \shl{\timeinseconds{1126}} & \timeinseconds{11872} & \hl{\timeinseconds{104035}} & \timeinseconds{1084} & \timeinseconds{11498} & \hl{\timeinseconds{101515}} \\
    & \arss & 1.4688 & 1.4643 & 1.4684 & \hl{0.006} & \hl{0.006} & \hl{0.005} & \timeinseconds{1363} & \timeinseconds{16564} & \timeinseconds{170035} & \timeinseconds{1120} & \timeinseconds{15913} & \timeinseconds{165872} \\
\midrule
\multirow{4}{*}{\setting{21}{H3}}
    & \ss   & 0.0694 & 0.1301 & 0.1511 & 0.653 & 0.349 & 0.245 & \hl{\timeinseconds{536}} & \shl{\timeinseconds{5266}} & \shl{\timeinseconds{32464}} & \hl{\timeinseconds{500}} & \shl{\timeinseconds{4980}} & \shl{\timeinseconds{30887}} \\
    & \ass  & 0.0551 & 0.0839 & 0.1218 & 0.725 & 0.581 & 0.391 & \timeinseconds{647} & \timeinseconds{5488} & \hl{\timeinseconds{20733}} & \timeinseconds{581} & \timeinseconds{5238} & \hl{\timeinseconds{19556}} \\
    & \rss  & 0.0629 & 0.0948 & 0.1447 & 0.686 & 0.526 & 0.277 & \shl{\timeinseconds{580}} & \hl{\timeinseconds{5261}} & \timeinseconds{48201} & \shl{\timeinseconds{512}} & \hl{\timeinseconds{4923}} & \timeinseconds{45668} \\
    & \arss & 0.1957 & 0.1971 & 0.1965 & \hl{0.036} & \hl{0.022} & \hl{0.021} & \timeinseconds{1204} & \timeinseconds{8759} & \timeinseconds{89965} & \timeinseconds{914} & \timeinseconds{8164} & \timeinseconds{86490} \\
\bottomrule
\end{tabular}
\caption{High-cost settings: accuracy and runtime across sample budgets.}
\label{tab:q2-2}
\end{table*}

\begin{figure*}[h!]
    \centering
    \begin{subfigure}[b]{0.32\textwidth}
        \centering
        \includegraphics[width=\textwidth]{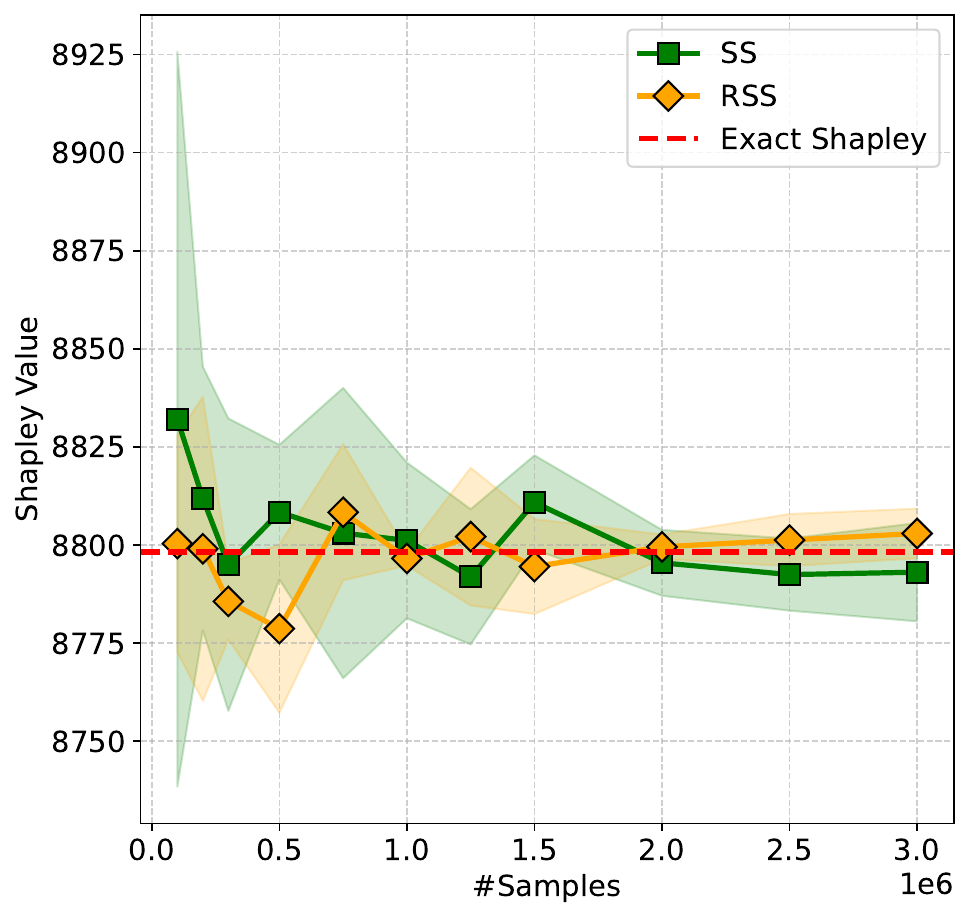}
        \caption{\rss vs \ss}
        \label{fig:10-14-7-1}
    \end{subfigure}
    \hfill
    \begin{subfigure}[b]{0.32\textwidth}
        \centering
        \includegraphics[width=\textwidth]{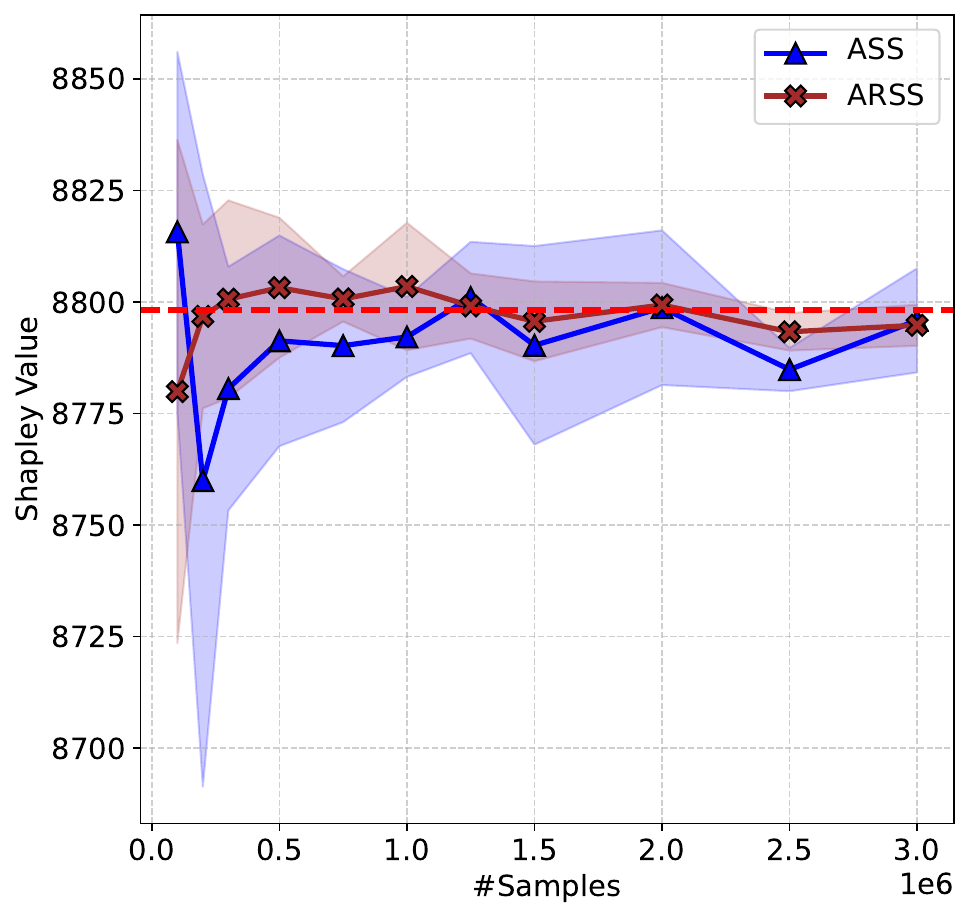}
        \caption{\arss vs \ass}
        \label{fig:10-14-7-2}
    \end{subfigure}
    \hfill
    \begin{subfigure}[b]{0.32\textwidth}
        \centering
        \includegraphics[width=\textwidth]{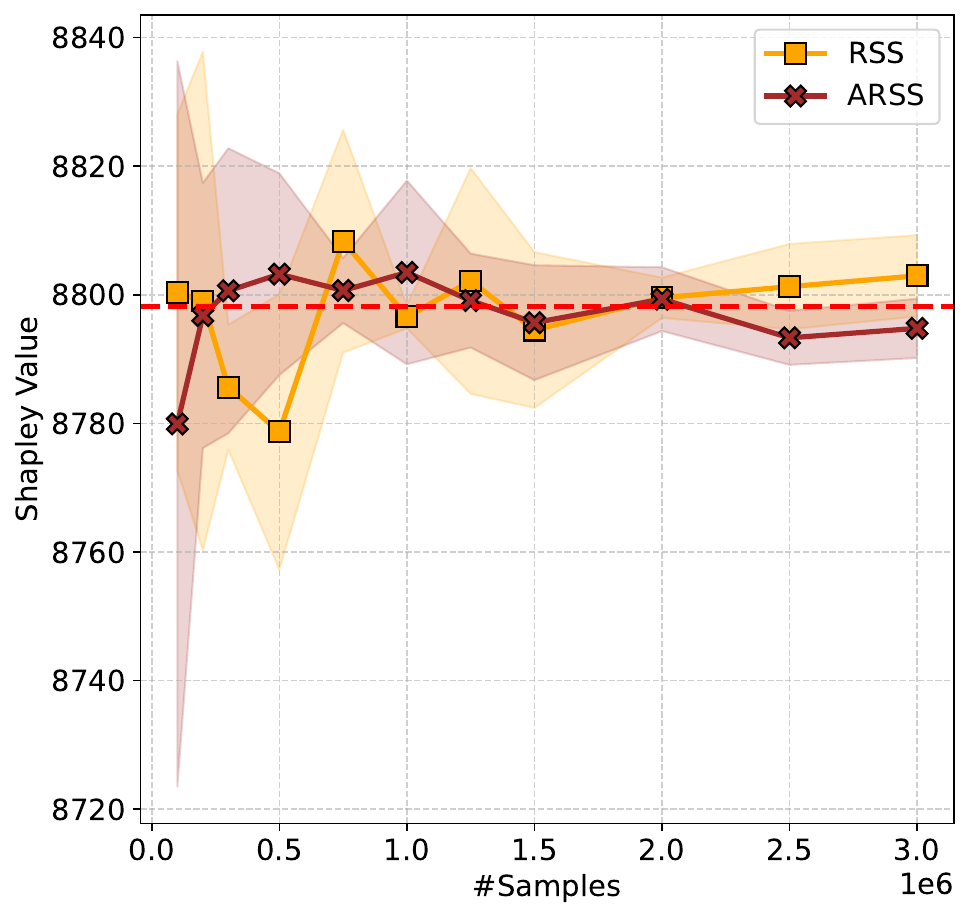}
        \caption{\arss vs \rss}
        \label{fig:10-14-7-3}
    \end{subfigure}
    \caption{Convergence comparisons for \setting{10}{L1}.}
    \label{fig:14-4}
\end{figure*}

\begin{figure}[h!]
    \centering
    \includegraphics[width=0.35\textwidth]{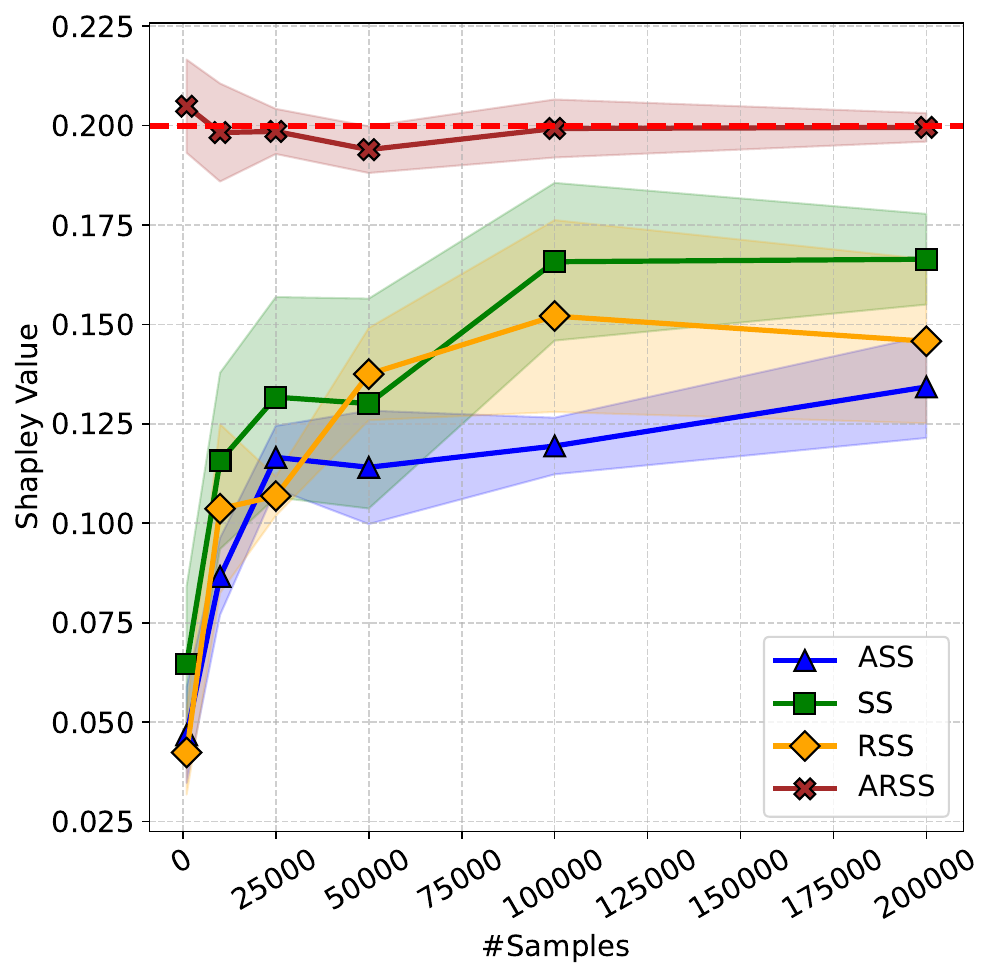}
    \caption{Convergence in high-cost setting \setting{21}{H3}.}
    \label{fig:21-h3-convergence}
\end{figure}

Convergence plots further clarify estimator dynamics. Figure~\ref{fig:14-4} shows pairwise comparisons for \setting{10}{L1}. \rss approaches the true value quickly and with less noise than \ss by reducing within-stratum variance through relation vectors. \arss improves further by concentrating budget on high-variance strata, yielding smoother and faster convergence than \ass, which adapts only over size-based strata. In high-cost \setting{21}{H3} (Figure~\ref{fig:21-h3-convergence}), early accuracy is crucial because each sample is expensive. \arss stabilizes rapidly with minimal samples and maintains the lowest \mre across budgets (\(0.036\) at 1k, \(0.022\) at 10k, \(0.021\) at 100k), whereas \ss and \rss converge more slowly and with higher variance.

\subsection{Optimizations}
\label{sec:opt}

Efficiency depends on limiting both the number of queries and the cost per query. Caching, static pruning, and indexing reduce redundant work in the naïve execution path. We additionally compile the original query \(Q\) into a reusable materialized view \(V_Q\) that encapsulates expensive joins and filters and a lighter query \(Q'\) that is parameterized by the sampled identifiers. During sampling, we evaluate variants of \(Q'\) over \(V_Q(D)\) rather than re-optimizing the full query for every coalition. This approach is analogous in spirit to knowledge compilation in logic~\cite{darwiche2002knowledge} and to classic view-based optimization and maintenance in databases~\cite{selinger1979access,gupta1995maintenance}. In our \textsc{TPC-H} experiments, precomputation enables substantial speedups for join-heavy queries and turns previously prohibitive cases into interactive-time evaluations.

Table~\ref{tab:tpch-runtime-new-time} summarizes original runtimes (\rtime), optimized runtimes (\texttt{new-time}), view build times (\texttt{mv-time}), and memory footprints (\texttt{mem-space}) for compiled views per \textsc{TPC-H} query. The optimization yields large improvements, particularly for queries with deep or selective joins, while leaving already light queries essentially unchanged. These results indicate that combining \rss/\arss with view compilation makes sampling-based Shapley estimation practical at interactive latencies.

\begin{table}[t!]
  \centering
  \begin{minipage}{\linewidth}
    \centering
    \begingroup
      \setlength{\tabcolsep}{4pt}
      \renewcommand{\arraystretch}{1.0}
      \begin{tabular}{@{}lcccc@{}}
        \toprule
        & \rtime & \texttt{new-time} & \texttt{mv-time} & \texttt{mem-space} \\
        \midrule
        \q{1}  & \formattime{1.6}{s}{1}   & \formattime{0.04}{s}{2}    & \formattime{0.16}{s}{2}    &  16 KB \\
        \q{2}  & \formattime{0.21}{s}{2}  & \formattime{0.12}{s}{2}    & \formattime{0.20}{s}{2}    &  264 KB \\
        \q{3}  & \formattime{0.63}{s}{2}  & \formattime{0.11}{s}{2}    & \formattime{1.9}{s}{1}     &  624 KB \\
        \q{4}  & \formattime{0.21}{s}{2}  & \formattime{0.2}{s}{1}     & \formattime{0.2}{s}{1}     &  189 MB \\
        \q{5}  & \formattime{0.28}{s}{2}  & \formattime{0.1}{s}{1}     & \formattime{0.3}{s}{1}     &  16 KB \\
        \q{6}  & \formattime{0.4}{s}{1}   & \formattime{0.04}{s}{2}    & \formattime{0.43}{s}{2}    &  15 MB \\
        \q{7}  & \formattime{0.36}{s}{2}  & \formattime{0.1}{s}{1}     & \formattime{0.6}{s}{1}     &  16 KB \\
        \q{8}  & \formattime{0.84}{s}{2}  & \formattime{0.19}{s}{2}    & \formattime{0.23}{s}{2}    &  4.1 MB \\
        \q{9}  & \formattime{1.67}{s}{2}  & \formattime{0.12}{s}{2}    & \formattime{0.97}{s}{2}    &  3.6 MB \\
        \q{10} & \formattime{0.36}{s}{2}  & \formattime{0.37}{s}{2}    & \formattime{0.88}{s}{2}    &  18 MB \\
        \q{11} & \formattime{0.09}{s}{2}  & \formattime{0.25}{s}{2}    & \formattime{0.22}{s}{2}    &  41 MB \\
        \q{12} & \formattime{0.33}{s}{2}  & \formattime{0.09}{s}{2}    & \formattime{0.77}{s}{2}    &  1 MB \\
        \q{13} & \formattime{1.56}{s}{2}  & \formattime{0.11}{s}{2}    & \formattime{1.17}{s}{2}    &  6.7 MB \\
        \q{14} & \formattime{0.22}{s}{2}  & 0.09 s                      & 0.62 s                      & 3.6 MB \\
        \q{15} & \formattime{0.36}{s}{2}  & 0.09 s                      & 0.6 s                       & 480 KB \\
        \q{16} & \formattime{0.65}{s}{2}  & 0.38 s                      & 0.27 s                      & 5.7 MB \\
        \q{17} & \formattime{31}{min}{0}  & 0.09 s                      & 4.5 s                       & 416 KB \\
        \q{18} & \formattime{3.6}{s}{1}   & 2.3 s                       & 7.6 s                       & 7.8 MB \\
        \q{19} & \formattime{0.58}{s}{2}  & 0.1 s                       & 2.1 s                       & 8.2 MB \\
        \q{20} & \formattime{70}{min}{0}  & 0.1 s                       & 2.2 s                       & 256 KB \\
        \q{21} & \formattime{0.6}{s}{1}   & 0.49 s                      & 1.04 s                      & 4.8 MB \\
        \q{22} & \formattime{0.61}{s}{2}  & 0.1 s                       & 0.27 s                      & 568 KB \\
        \bottomrule
      \end{tabular}
    \endgroup
    \caption{Runtime (\rtime), optimized runtime (\texttt{new-time}), view build time (\texttt{mv-time}), and memory footprint (\texttt{mem-space}) for compiled views across \textsc{TPC-H} queries.}
    \label{tab:tpch-runtime-new-time}
  \end{minipage}
\end{table}

\subsection{Discussion and Takeaways}

Relation-aware stratification yields tangible benefits for Shapley estimation. By organizing coalitions according to relation vectors, \rss reduces intra-stratum variance, improves sample efficiency, and often collapses variance entirely for certain queries, enabling exact recovery with few samples. Adaptive allocation then amplifies these gains by steering budget toward strata that remain variable, which is critical in heterogeneous relational workloads. The combination in \arss consistently lowers \mre, accelerates convergence, and reduces wasted query evaluations relative to size-based stratification. While managing many strata introduces some overhead, our view-based compilation and caching reduce per-sample costs, and the variance-driven savings in \qtime typically dominate. Limitations include the growth of the stratification space for many-relation queries and the memory footprint of per-stratum statistics at very high budgets. Coarsening relation vectors by quantiles and pruning structurally unproductive vectors mitigate these issues in practice.

Overall, the experiments indicate that \arss provides a practical and accurate approach to Shapley estimation for relational queries under realistic costs. It offers early, high-quality estimates and stable convergence profiles, making it suitable for interactive explanation workflows where query evaluations are expensive and time is constrained.

\section{Conclusion and Future Work} 
\label{sec:conclusion}

This chapter aims to review the challenges solved in the thesis and summarize all of the discussions we have had so far, and also highlight potential areas for future research in the domain of relation stratified sampling. While many existing works have explored various sampling approaches—most notably Monte Carlo and stratified sampling—these methods typically overlook the structural aspects of the query, such as the underlying relational schema or the roles of different relations involved in the computation. In this work, we address this gap by proposing a new sampling approach that takes into account the number of tuples drawn from each relation and stratifies samples accordingly. This relation-aware stratification allows us to more effectively capture the structural diversity inherent in query provenance.

Our proposed algorithm explores all possible combinations of contributing tuples across the relations involved in a query. In practice, this results in a potentially exponential number of subsets, particularly as the size of the database grows. To address the computational overhead this introduces, we propose a hashing mechanism to store previously seen subsets along with their computed marginal contributions. This allows us to avoid redundant evaluations and significantly reduce query processing time during sampling. The effectiveness of our proposed methods has been demonstrated in the preceding chapter. Our experimental results show that relation-stratified sampling consistently outperforms both baseline approaches—Monte Carlo and traditional stratified sampling—in terms of convergence to the true Shapley value and reduction of estimation variance.
 Now, we wish to outline some potential future research directions:

\subsection{Query Compilation via Materialized Views} 
\label{sec:future-q-compile}

An interesting direction for future work involves accelerating Shapley value estimation through query compilation, particularly by leveraging materialized views. The core idea would be to precompute and cache intermediate results that are shared across many of the marginal queries evaluated during sampling. This would help avoid repeated computation of expensive operations, especially in large-scale settings with high query reuse.

In Shapley estimation, many marginal queries share a common core structure and differ only in a few predicates or tuple-level selections. For example, each marginal contribution query typically applies the same join conditions and selection predicates over different subsets of the input. This structural similarity opens the door to defining a materialized view that captures the static part of the computation—such as common joins and filters—which can then be reused across all sampling queries.

Once defined, this view could serve as a compiled query representation, enabling subsequent marginal queries to operate on a smaller, pre-filtered dataset instead of scanning the full base relations. This has the potential to significantly reduce runtime by pushing the cost of expensive operations like multi-way joins into a one-time setup step.

Rewriting each sampling query to refer to the materialized view instead of the full base tables would preserve correctness while benefiting from faster evaluation. In principle, this transformation acts as a domain-specific form of knowledge compilation: converting the original parameterized query into a more efficient representation tailored to repeated evaluation.

We expect that such an approach could lead to substantial runtime improvements, potentially achieving order-of-magnitude speedups in settings where the number of marginal queries is large and the overlap between them is high. Moreover, the overhead of constructing the view would be incurred only once, and the memory cost would remain modest due to filtering on relevant subsets.

This line of optimization is complementary to the hash-based caching strategy discussed earlier. While caching helps eliminate redundant execution of identical queries, materialized view compilation could reduce the cost of evaluating structurally similar—but not identical—queries. Investigating this hybrid query processing strategy in future work may offer a scalable path forward for efficient Shapley value estimation over large datasets.

\subsection{Quantile-Based Stratification for Scalable Sampling} 
\label{sec:future-quantile}

One limitation of relation-stratified sampling, as currently designed, is the rapid growth in the number of strata. In queries involving multiple relations, each stratum is defined by a size vector $(s_1, \dots, s_r)$, where $s_j$ represents the number of tuples drawn from relation $R_j$. While this fine-grained stratification captures coalition structure with high fidelity, it also leads to an exponential increase in the number of strata when relations contain many endogenous tuples. As a result, the sampling budget may be spread too thin, limiting efficiency and increasing the risk of under-sampling high-variance regions.

A promising direction for future work is to explore \emph{quantile-based stratification} as a scalable alternative. Instead of stratifying based on exact tuple counts, each relation could be partitioned into quantile intervals—such as quartiles or deciles—based on relevant tuple attributes or statistical properties. Strata would then be indexed by quantile vectors $(q_1, \dots, q_r)$, where each $q_j$ refers to the quantile bin sampled from relation $R_j$.

This coarser stratification scheme would substantially reduce the total number of strata, replacing an exponential dependence on tuple counts with a more tractable cross-product of quantile levels. At the same time, it would preserve relation-aware structure and still allow for meaningful differentiation between coalition types. Such an approach offers a tunable balance between stratification granularity and computational feasibility, making it well-suited for large-scale settings.

Quantile-based stratification could also be integrated with adaptive sampling strategies—such as dynamic Neyman allocation—to prioritize strata with higher observed variance. This combination may offer an effective trade-off between scalability and estimation accuracy, enabling efficient Shapley value approximation even in complex or high-dimensional datasets.

We consider this a valuable avenue for future exploration, particularly for systems aiming to scale relation-aware Shapley estimation to industrial or web-scale databases.

\bibliographystyle{ieeetr}
\bibliography{ref}
\end{document}